\documentclass[3p,times,review]{elsarticle}

\usepackage[utf8]{inputenc}
\usepackage[english]{babel}
\usepackage{amsmath}
\usepackage{amsfonts}
\usepackage{mathrsfs}
\usepackage{amssymb}
\usepackage{xfrac}
\usepackage{graphicx}
\usepackage[export]{adjustbox}
\usepackage{enumitem}
\usepackage{booktabs}
\usepackage{hyperref}
\usepackage{setspace}
\usepackage{multirow}
\usepackage{lineno}
\usepackage{array}
\usepackage{lipsum}

\usepackage{xcolor}
\hypersetup{
	colorlinks,
	linkcolor={black},
	citecolor={blue!50!black},
	urlcolor={blue!80!black}
}

\newcolumntype{C}[1]{>{\centering\arraybackslash}m{#1}}
\newcolumntype{L}[1]{>{\arraybackslash}m{#1}}

\AtBeginDocument{
	\heavyrulewidth=.08em
	\lightrulewidth=.05em
	\cmidrulewidth=.03em
	\belowrulesep=.65ex
	\belowbottomsep=0pt
	\aboverulesep=.4ex
	\abovetopsep=0pt
	\cmidrulesep=\doublerulesep
	\cmidrulekern=.5em
	\defaultaddspace=.5em
}

\makeatletter
\def\@author#1{\g@addto@macro\elsauthors{\normalsize%
    \def\baselinestretch{1}%
    \upshape\authorsep#1\unskip\textsuperscript{%
      \ifx\@fnmark\@empty\else\unskip\sep\@fnmark\let\sep=,\fi
      \ifx\@corref\@empty\else\unskip\sep\@corref\let\sep=,\fi
    }%
    \def\authorsep{\unskip,\space}%
    \global\let\@fnmark\@empty
    \global\let\@corref\@empty  
  \global\let\sep\@empty}%
  \@eadauthor={#1}
}
\makeatother

\journal{Nuclear Instruments and Methods in Research A}

\begin{document}
\begin{frontmatter}
  \author[PHY,DREXEL]{T.~Polakovic}
\author[PHY]{W.R.~Armstrong}
\author[HEP]{V.~Yefremenko}
\author[MSD]{J.E.~Pearson}
\author[PHY]{K.~Hafidi}
\author[DREXEL,DREXEL2]{G.~Karapetrov}
\author[PHY]{Z.-E.~Meziani}
\author[MSD]{V.~Novosad\corref{cor1}}
\ead{novosad@anl.gov}

\address[PHY]{Physics Division, Argonne National Laboratory, Argonne, IL}
\address[HEP]{High Energy Physics Division, Argonne National Laboratory, Argonne, IL}
\address[MSD]{Materials Science Division, Argonne National Laboratory, Argonne, IL}
\address[DREXEL]{Department of Physics, Drexel University, Philadelphia, PA}
\address[DREXEL2]{Department of Materials Science and Engineering, Drexel University, Philadelphia, PA}

\cortext[cor1]{Corresponding author}

  \title{Superconducting nanowires as high-rate photon detectors in strong magnetic fields} 		

  \begin{abstract}
   Superconducting nanowire single photon detectors are capable of single-photon detection across
   a large spectral range, with 
   near unity detection efficiency, picosecond timing jitter,  and sub-10~$\mu$m position resolution
   at rates as high as 10$^{9}$~counts/s. In an effort to bring this technology into nuclear physics experiments,
   we fabricate Niobium Nitride nanowire detectors using ion beam assisted sputtering and
   test their performance in strong magnetic fields. We demonstrate that these devices are capable of
   detection of 400~nm wavelength photons with saturated internal quantum efficiency at temperatures of 3 K and in magnetic fields potentially up to 5~T at high rates and with nearly zero dark counts.
   \end{abstract}

\end{frontmatter}


\section{Introduction}

 Superconducting nanowire single photon detectors (SNSPD) are a relatively recent technology~\cite{gol2001picosecond}
 that shows great promise due to their detection capabilities that are in many aspects superior
 to more conventional semiconductor detectors: timing jitter (FWHM of the distribution of deviation from an ideal periodic
 single-photon-response) of $\lesssim$~15~ps~\cite{wu2017improving}, near-unity
 detection efficiency~\cite{marsili2013detecting}, and count rate higher than 10$^{9}$~counts/s with 10$^{-3}$~count/s dark counts~\cite{shibata2015ultimate}.
 These metrics make SNSPDs a popular choice in fields of quantum communication and sensing, where they have been
 used in quantum key distribution~\cite{takesue2007quantum}, long-range quantum
 teleportation experiments~\cite{takesue2015quantum,valivarthi2016quantum}, or LIDAR systems~\cite{zhu2017demonstration}.
 While SNSPDs are inherently a broadband detector and there are efforts to fabricate efficient devices for use in the UV and visible 
 range~\cite{wollman2017uv,slichter2017uv}, most of the mentioned applications work with standardized IR telecom wavelengths, so the
 detector development is focused on optimization of detection efficiency of low energy
 photons~\cite{marsili2013detecting, bellei2016free, marsili2013mid, csete2011numerical, wang2016broadband}.
 The situation is, however, different if one would want to use SNSPDs for experiments in nuclear physics, where potential
 applications would include detection of Cherenkov radiation, light from ionization or from scintillator, and active
 polarized targets~\cite{biroth2016design}, where the spectral density is shifted towards visible-UV range~\cite{nikl2013development}
 or as a part of detectors where it can be utilized for direct detection of $\alpha$-- and $\beta$--particles~\cite{azzouz2012efficient} or
 electrons~\cite{rosticher2010high}.
 Because SNSPDs have a trivial footprint, they can be positioned closer to the active area of the experiment and
 in this case we need to focus on performance in conditions that are typically not seen in experiments related to quantum communication.
 The complications associated with these environments are primarily large magnetic fields~\cite{armstrong2015threshold} and, often times,
 liquid helium temperatures~\cite{biroth2016design}, where Si-based detectors are known to underperform~\cite{biroth2015silicon}. As SNSPDs are
 superconducting detectors, cryogenic environments do not degrade their performance. On the other
 hand, their detection capabilities in strong magnetic fields have not been extensively studied, and so far,
 SNSPD characterization has usually been limited to fields smaller than 
 0.2~T~\cite{korneev2014characterization,engel2012dependence,lusche2012effect}.
 
 In this work we explore the detection capabilities of 400~nm wavelength photons in high magnetic fields and we show that, by
 using the recently developed ion beam assisted sputtering~\cite{polakovic2018room}, we can fabricate Niobium Nitride
 (NbN) high-rate, low dark count SNSPDs capable of operation at 3 K and in magnetic
 fields as high as 5~T -- a dramatic performance increase compared to Si-based
 detectors~\cite{biroth2015silicon,xie2018rate}.
 
\section{Device Fabrication}

\begin{figure}[htb]
  \centering
  \includegraphics[width=0.45\textwidth]{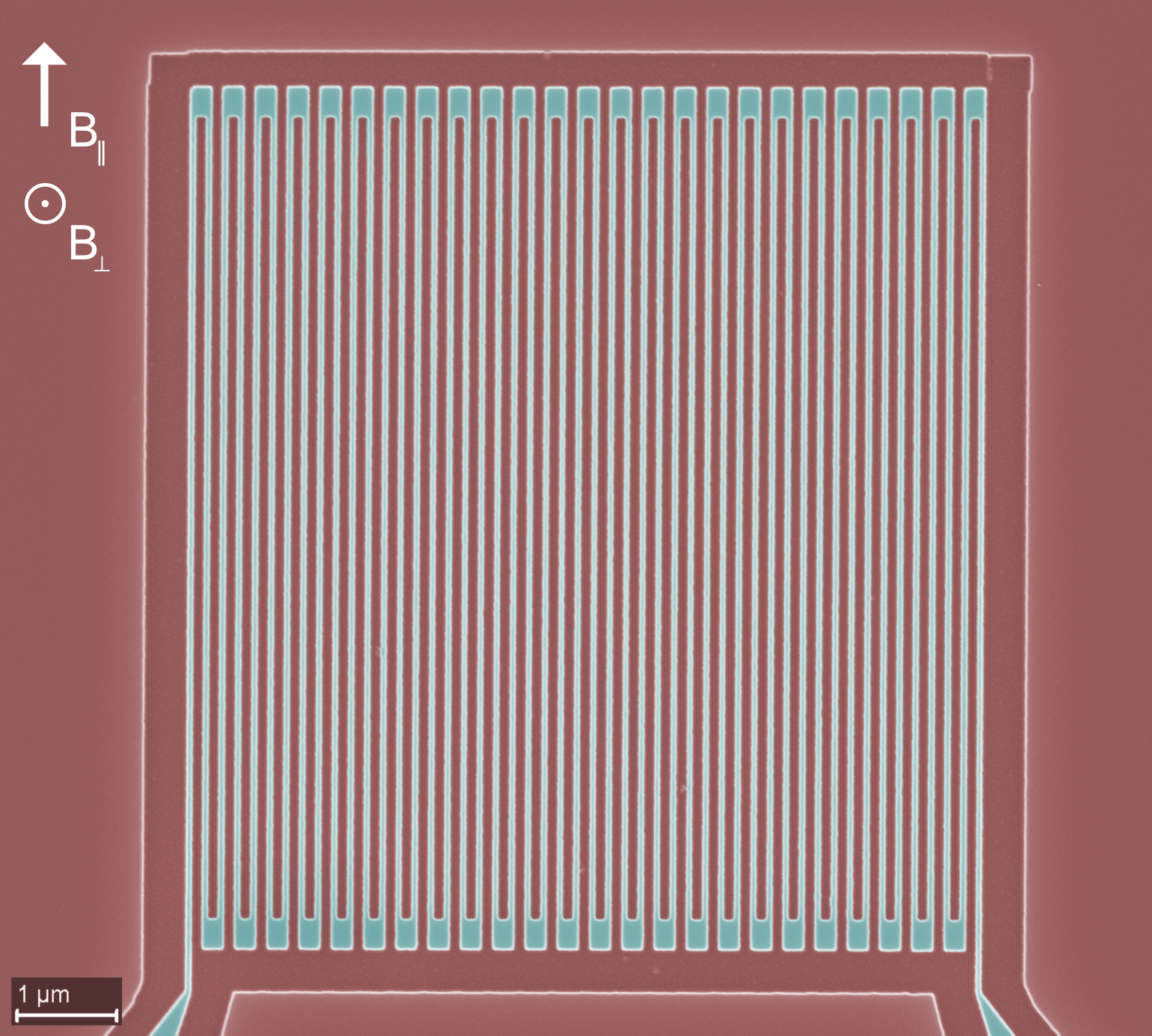}
  \caption{False color SEM micrograph of the fabricated nanowire detector, where the active current-carrying device is colored in teal.
  Field directions used in this experiment are depicted in the top-left corner. The voltage is sampled at the vicinity of the two points where
  the meander is connected to the external wiring.}
  \label{fig:wire_sem}
\end{figure}

 The detectors used in this work have the standard meander geometry, with a wire thickness of 13.5~nm, wire width of 80~nm,
 spacing between the wires of 110 nm and a pixel size of 10~$\times$~10~$\mu$m$^2$ (as shown in Figure~\ref{fig:wire_sem}).
 The stoichiometric NbN thin films were prepared by ion beam assisted sputtering~\cite{polakovic2018room} at room temperature,
 with Ar as sputtering gas at 2~$\times$~10$^{-3}$~torr in a ultra-high vacuum sputtering system from Angstrom
 engineering~\cite{Angstrom}. Devices were patterned using electron beam lithography and ZEP~520A diluted at 1:2 with anisol
 as resist. Nanowires were etched by reactive ion etching in CF$_4$ plasma.

 Films before patterning had a superconducting critical temperature $T_C$~=~8~K
 and normal sheet resistance of approximately 683~$\Omega$. The perpendicular critical magnetic field was
 determined to be H$_{C2}$(0)~=~32~T and the coherence length $\xi$(0)~=~3.2~nm~\cite{polakovic2018room}.
 After patterning, the nanowire detector's $T_C$ remained unchanged and the critical current density was determined to be
 $j_C$~=~2.2$\times$10$^{10}$~A/m$^2$ using voltage criterion of 2~$\mu$V across the total length of the meander (approximately
 700 $\mu$m).

\section{Experimental setup}

 A Quantum Design Physical Property Measurement System (PPMS) was used to control temperature and apply magnetic field during
 measurement. The characterization
 apparatus consisted of a custom designed PPMS insert manufactured by Quantum Opus LLC coupled with a Opus~One SNSPD
 bias and readout module~\cite{QOpus} and R\&S~RTM3000 oscilloscope. The signal was measured using a two-point voltage readout.
 Light to the detectors was supplied from flood illumination by InGaN LED integrated into the PPMS insert. Nominal wavelength of the
 LED was 465~nm at room temperature and, when cooled to operational temperature, the wavelength blueshifted to approximately
 400~nm. Unless otherwise specified, the LED forward bias was set to 30~$\mu$A to minimize excess device heating.
 All measurements were conducted at temperatures of 3~K due to better temperature stability. We observe negligible change in
 performance at 4~K. All measurements in magnetic fields were carried out by zero-field cooling to 3~K before applying magnetic
 fields.

\section{Results}

\begin{figure}[htb]
  \centering
  \includegraphics[width=0.5\textwidth,trim= 10 12 0 13,clip]{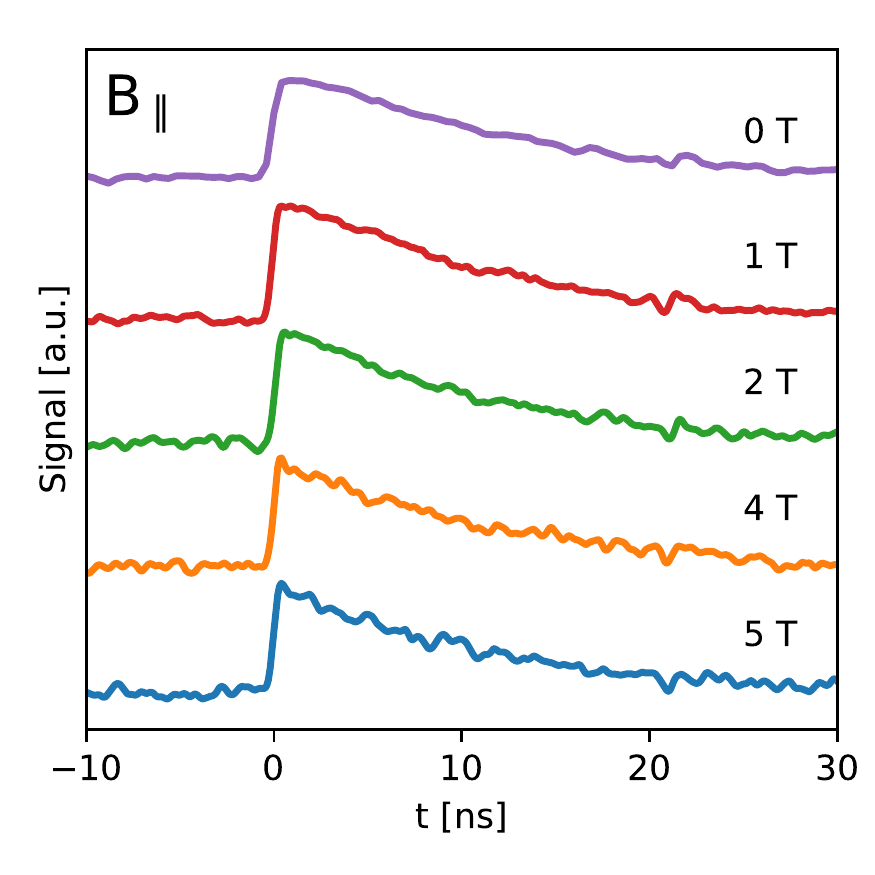}
  \caption{Waterfall plot of typical single-photon voltage pulse waveforms at various parallel magnetic fields. All signals are normalized
  to their respective pulse maximum and signal baseline can be seen at negative times.}
  \label{fig:pulses}
\end{figure}

 In this work, we explore the device performance in two field configurations: One is in field applied parallel to the device plane
 (B$_{\parallel}$) and one with field perpendicular to the device plane (B$_{\perp}$) as schematically shown in
 Figure~\ref{fig:wire_sem}. Typical time trace of photon detection events can be seen in Figure~\ref{fig:pulses}.
 The 20-80\% rise time was determined to be $\tau_R$~=~341~$\pm$~31~ps and 80-20\% fall time is $\tau_F$~=~11.78~$\pm$~1.6~ns.
 These values didn't change significantly as a function of applied field or light intensity.
 
 The important detection characteristics of a SNSPD device can be extracted from the dependence of count rate as a function of device
 bias current. At low currents, the probability of quasi-particle excitation and formation of a hot-spot region after photon absorption is
 low~\cite{natarajan2012superconducting,marsili2016hotspot} and increases with increasing bias current. As one increases the constant
 current bias of a device further, the count rate reaches a plateau - the saturated internal
 efficiency~\cite{baek2011superconducting,najafi2014fabrication}, where the probability of detecting an absorbed photon is close to
 unity~\cite{lusche2012effect,marsili2011single}. At these current values, the total detection efficiency is determined by external
 parameters such as geometric filling factors~\cite{yang2009suppressed} or optical coupling~\cite{hu2009efficiently}.

\begin{figure}[htb]
  \centering
  \includegraphics[width=0.5\textwidth,trim= 10 16 0 0,clip]{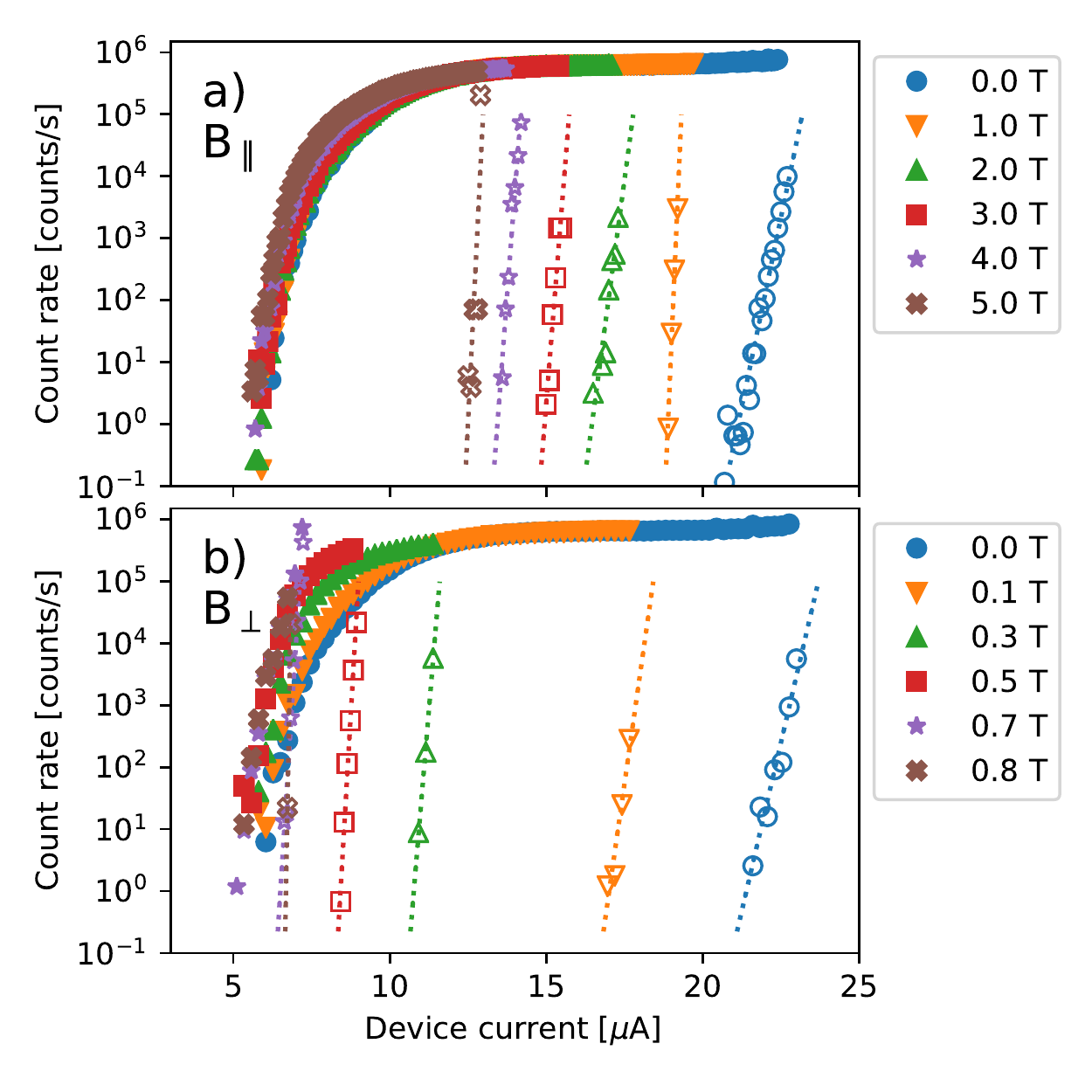}
  \caption{Dependence of count rate as a function of nanowire bias current at various parallel magnetic fields at constant LED illumination intensity.
  Total counts are plotted with full circles, dark counts with empty circles. Dotted lines are exponential fits to the dark count rate data.
  Top and bottom figure corresponds parallel and perpendicular fields,
  respectively.}
  \label{fig:RvI}
\end{figure}

\begin{figure}[htb]
  \centering
  \includegraphics[width=0.5\textwidth,trim= 7 5 0 0,clip]{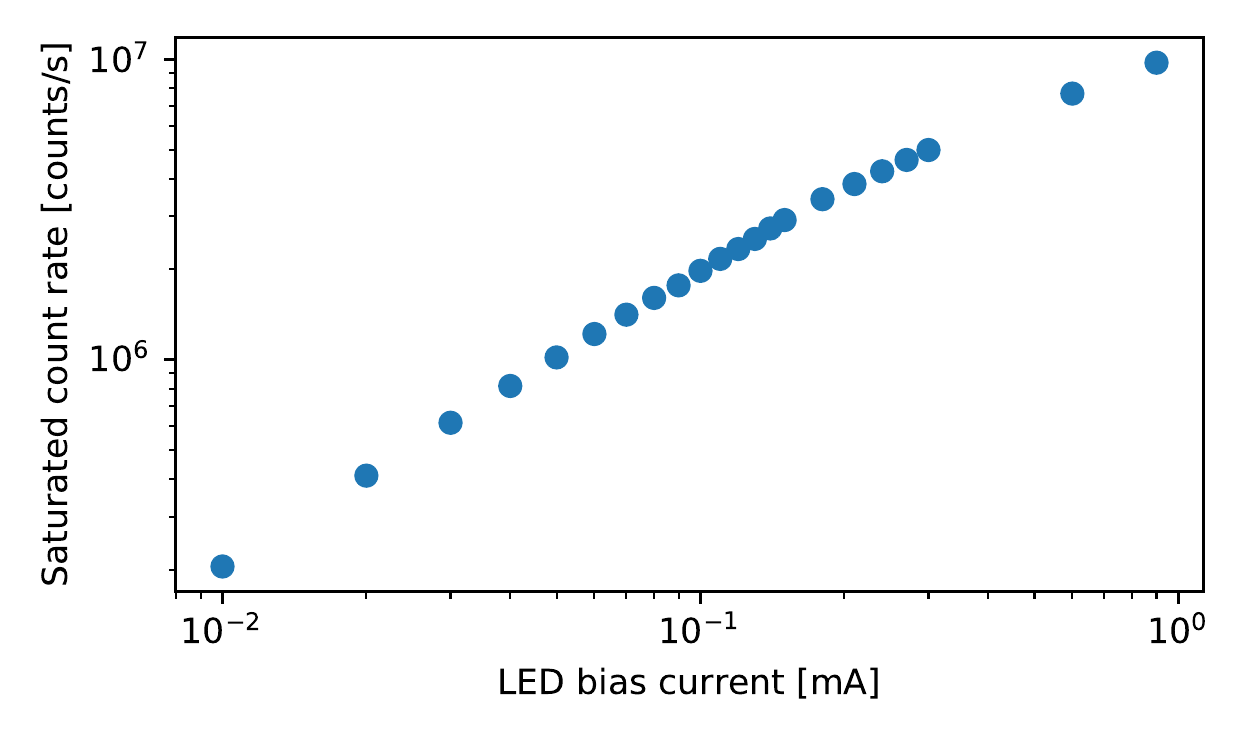}
  \caption{Saturated count rate as a function of 400 nm LED forward bias current. In this small current forward bias
  regime, the light intensity is proportional to the InGaN LED current.}
  \label{fig:SRvLED}
\end{figure}

 For 400~nm photons, in zero magnetic field, our devices can reach saturated internal efficiency at bias currents of
 approximately 9~$\mu$A, well below the critical currents of 23~$\mu$A (current density close to j~=~2.2$\times$10$^{10}$~A/m$^2$).
 This means that the devices are capable
 of high detection rate, with zero dark counts which increase exponentially as the bias current reaches the critical
 value (as can be seen in Figure~\ref{fig:RvI}). The largest observed count rate achieved with our devices was
 approximately 10$^{7}$~counts/s for the 100~$\mu\mathrm{m}^2$ device. The device count rate up to these values is linear with increased
 photon flux (which is proportional to the LED forward bias current), which confirms that we operate in the single-photon
 regime~\cite{marsili2011single}. This is not the maximum count rate capability of the
 SNSPD device (as can be seen from the trend in Figure~\ref{fig:SRvLED}), but a limit imposed by the temperature
 control capabilities of our setup, where the heat load of the LED exceeds the cooling power of the PPMS. The
 deviation from the expected linear trend in Figure~\ref{fig:SRvLED} is attributed to these heating effects. In absence
 of this spurious heating, we expect our SNSPD devices to be capable of detection rates of the order of 10$^{8}$~counts/s -
 determined by fall time, where count rate is proportional to $1/\tau_F$~\cite{kerman2013readout}. If one desires
 to achieve higher count rates, common approaches include decreasing the length of the wire (to decrease the
 kinetic inductance of the device $L_K$~\cite{kerman2006kinetic}) or introduce a shunt resistance $R_S$ to decrease
 the fall time constant $\tau_F \propto L_K / R_S$ or to split the wire into multiple segments connected in parallel
 so that the total inductance is a harmonic mean of the individual segment 
 inductances~\cite{ejrnaes2009characterization,heath2014nano}.

\begin{figure}[htb]
  \centering
  \includegraphics[width=0.5\textwidth,trim= 0 15 0 10, clip]{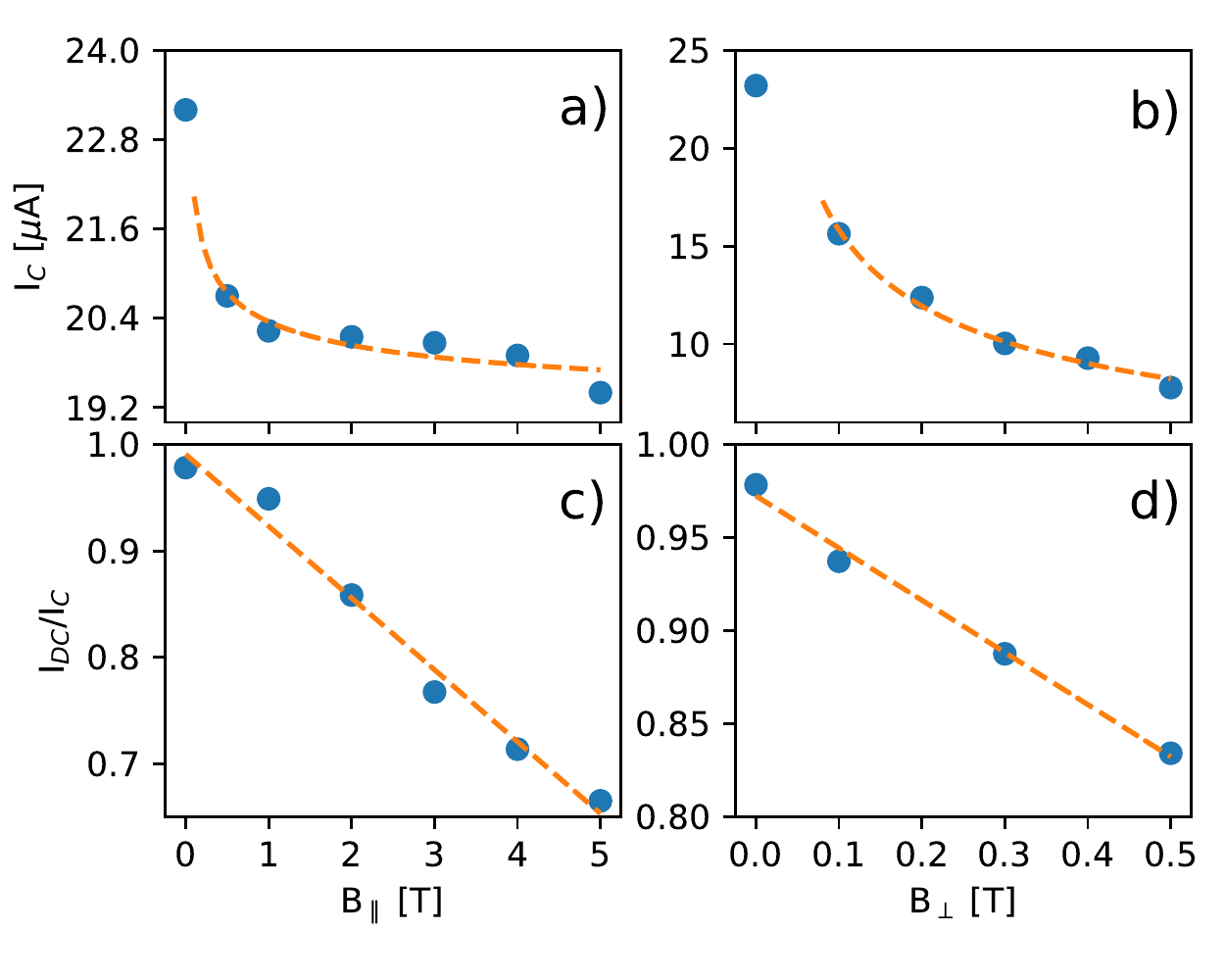}
  \caption{Dependence of superconducting critical currents I$_C$ (top row) and the normalized detector
  critical currents I$_{DC}$/I$_C$ (bottom row) on applied magnetic fields. Left and right column
  corresponds to parallel and perpendicular field orientations, respectively. Dashed lines in are fits of
  linear dependence in plots c) and d), and B$^{-\alpha}$ dependence in a) and b).}
  \label{fig:ICs}
\end{figure}

 Before we get to detector performance in applied magnetic fields, we'll briefly discuss the superconducting
 critical currents of the nanowires without LED illumination. The observed power law dependence of superconducting critical
 currents I$_C$ on magnetic fields can be explained by the vortex dynamics of a type--II superconductor
 with strong edge pinning where the current should be inversely proportional to the applied
 field~\cite{maksimova1998mixed}:
 
 \begin{equation}
 	j_C(B) = j_C (0) \cdot \frac{\mathcal{B}}{2B},
 \end{equation}
 where $\mathcal{B} = \frac{\Phi_0}{2\pi} \frac{1}{W\xi _{GL}}$ is the vortex penetration field, with $\Phi_0$ being the flux quantum,
 W the wire width and $\xi_{GL}$ the Ginzburg-Landau coherence length. However, our data
 doesn't fit a functional form proportional to B$^{-\alpha}$ with
 $\alpha \approx$~1, but rather $\alpha$ = 0.4 for perpendicular fields (Figure~\ref{fig:ICs}b) and
 $\alpha$ = 0.02 in case of fields parallel to the meander (Figure~\ref{fig:ICs}a). The case of perpendicular fields,
 with $\alpha \approx$ 0.5 has been observed in similar superconducting nanostructures before and can be explained
 by strong bulk vortex pinning~\cite{vodolazov2009strong,ilin2014magnetic} -- as one would anticipate from materials with
 high density of meso-scale lattice defects like our films grown by ion beam assisted sputtering. In the case of magnetic
 fields parallel to the transport current, the Lorenz force acting on the vortex lines is effectively zero and the critical current
 density becomes a function of number of vortices in the wire and their
 interactions~\cite{stejic1994effect}, which coupled with geometry-enhanced surface pinning
 greatly increases the critical current density~\cite{carneiro1998equilibrium}, as can be seen in our measurements.
 
 In the magnetic field dependence of the detector response, we will first discuss the case of magnetic field applied
 perpendicular to the device plane (see Figure~\ref{fig:RvI}b) which can be compared to results in
 literature~\cite{korneev2014characterization,engel2012dependence,lusche2012effect}. We can see a relatively strong
 strong deterioration of detection capabilities even at fields smaller than 1~T. This can be explained by the
 dynamics of the supercurrents~\cite{hortensius2012critical} and superconducting
 vortices~\cite{bulaevskii2012vortex} in external magnetic field. 
 Our choice of fabricating these devices out of NbN prepared by ion beam
 assisted sputtering, which has a comparatively higher values of critical field $H_{C2}$ and critical current densities\cite{polakovic2018room},
 has already led to a considerable improvement in field performance {olive}(by a factor of 2) when compared to comparable devices studied in
 literature~\cite{korneev2014characterization,engel2012dependence,lusche2012effect}, and achieves results
 comparable to devices with geometry optimized for performance in magnetic fields~\cite{charaev2016enhancement}.
 The practical limit of external
 perpendicular magnetic field is around 0.5~T, past which the detection is dominated by dark counts arising from
 fluctuations of the near-critical superconducting state~\cite{yamashita2011origin}. One can increase this value by
 engineering a stronger thermal coupling to the heat sink (i.e. the substrate)~\cite{hofherr2012dark}, introducing
 stronger vortex pinning centers to minimize vortex creep and vortex hopping~\cite{larkin1979pinning} optimizing the
 device geometry to prevent current crowding at the meander turns~\cite{charaev2016enhancement, akhlaghi2012reduced} or by increasing the
 wire cross-section. The last two methods, however, lead to a decrease in total detection efficiency by sacrificing the
 geometric filling ratio or the hot-spot expansion probability, respectively.
 
 As many experimental setups in nuclear physics are axially symmetric, with magnetic field applied along the symmetry
 axis (e.g. central solenoids in particle collider detectors~\cite{herve2000cms,lighthall2010commissioning}), it is also
 important to explore the behavior of the SNSPD devices in external fields aligned parallel to the detector plane.
 The quantitative dependence of rate as a function of bias current in parallel magnetic fields is different, as can
 be seen in the results plotted in Figure~\ref{fig:RvI}a. The
 detector reaches internal efficiency saturation in parallel magnetic fields as high as 5~T (the highest
 field achievable with our experimental setup), with
 device saturating at approximately 10~$\mu$A, well below the onset of dark counts at 12.5~$\mu$A. By extending the trend
 in Figure~\ref{fig:ICs}c, we can make a conservative estimate of the limiting parallel magnetic field which we believe to be
 approximately 8 T, if we assume that the saturation current is independent of the applied magnetic field. The assumption
 of constant saturation current doesn't necessarily hold, as can be seen in Figure~\ref{fig:RvI}, where the onset of
 saturation happens at lower bias currents. This behavior is assumed to be due to effects of magnetic vortices, which
 can assist the hotspot formation and expansion~\cite{bulaevskii2012vortex}. While there exist experimental studies of
 this effect perpendicular fields~\cite{vodolazov2015vortex}, the dynamics of the detection process in parallel fields
 might warrant a separate study, especially in the context of detection of IR photons, which is not the focus of this work.
 It is important to mention that these effects are relatively weak and lead to underestimation of the limiting magnetic
 field, so we believe that the value of 8 T is still a reasonable approximation, even if we're unable to reach magnetic fields of
 such magnitudes.
 
 As the physics driving this behavior is similar to the situation in perpendicular magnetic fields, one can use similar
 approaches to increase the value of critical magnetic fields: Increase the wire thickness and thermal coupling, change
 material microstructure to introduce additional vortex pinning sites, or optimize the meander turn geometry to
 decrease the current densities.
 
\section{Conclusion}

 We have demonstrated that superconducting nanowire single photon detectors are a viable technology for detection
 of individual photons in strong magnetic fields. We show that detectors fabricated from NbN prepared by ion beam
 assisted sputtering can withstand strong magnetic fields as high as 5~T in certain configurations and 0.5 T in the
 cases reported in literature, which is double the commonly reported field strength. Even at such large fields
 they are capable of high-rate detection of 400~nm photons (potentially up to 10$^{8}$~count/s at 100~$\mu$m$^2$ pixel size)
 with less than 1 dark counts per second, which makes them an attractive alternative for high-rate, low-background
 measurements in strong field environments - a common demand in nuclear physics experiments that cannot be met
 by conventional semiconductor detectors.

\section*{Acknowledgements}
We would like to thank Aaron Miller from Quantum Opus LLC for fruitful discussions and assistance in detector
characterization setup.

%
%
%
This work was supported by the U. S. Department of Energy (DOE), Office of Science, Offices of Nuclear Physics,
Basic Energy Sciences, Materials Sciences and Engineering Division under Contract \# DE-AC02-06CH11357.
A portion of this work was conducted at the Center for Nanoscale Materials, a U.S. Department of Energy, Office
of Science (DOE-OS) user facility.
\section*{References}


\bibliographystyle{elsarticle-num}

\end{document}